\documentstyle[12pt]{article}
\newcommand {\beq}{\begin{equation}}
\newcommand {\eeq}{\end{equation}}
\newcommand {\beqa}{\begin{eqnarray}}
\newcommand {\eeqa}{\end{eqnarray}}
\newcommand {\n}{\nonumber \\}

\begin{document}
\setlength{\oddsidemargin}{0cm}
\setlength{\baselineskip}{7mm}

\begin{titlepage}
 \renewcommand{\thefootnote}{\fnsymbol{footnote}}
$\mbox{ }$
\begin{flushright}
\begin{tabular}{l}
KEK-TH-799 \\
Jan. 2002
\end{tabular}
\end{flushright}

~~\\
~~\\
~~\\

\vspace*{0cm}
    \begin{Large}
       \vspace{2cm}
       \begin{center}
         {Vertex Operators in IIB Matrix Model}      \\
       \end{center}
    \end{Large}

  \vspace{1cm}

\begin{center}
           Yoshihisa K{\sc itazawa}\footnote
           {
e-mail address : kitazawa@post.kek.jp}

         {\it Institute of Particle and Nuclear Studies,\\
         High Energy Accelerator Research Organization (KEK),}\\
               {\it Tsukuba, Ibaraki 305-0801, Japan} \\
\end{center}

\vfill

\begin{abstract}
\noindent
\end{abstract}
The vertex operators for the supergravity multiplet
can be constructed through the Wilson lines in IIB matrix model.
We  investigate the structure of the vertex operators
and the symmetry of their correlation functions.
For this purpose,
we perturb the theory by the Wilson lines dual to
the supergravity multiplet.
The structure of the Wilson lines can be determined by
requiring $\cal{N}$=2 SUSY under the low energy approximation.
We argue that the generating functional of the correlators is invariant
under local SUSY transformations of IIB supergravity.

\vfill
\end{titlepage}
\vfil\eject

\section{Introduction}
\setcounter{equation}{0}
One of the important problems in matrix model formulations
of superstring theory is to identify gauge symmetries
which are expected in closed string theory\cite{BFFS}\cite{IKKT}.
In particular, we must identify local symmetries
of supergravity.
In the case of IIB matrix model, we expect to find local
supersymmetry, local Lorentz, general coordinate transformation
and three other gauge symmetries.
Since the rest follows from local supersymmetry, it is
expected to shed light on other gauge symmetries as well.

We recall IIB matrix model action \cite{IKKT}:
\beq
S  =  -{1\over g^2}Tr({1\over 4}[A_{\mu},A_{\nu}][A^{\mu},A^{\nu}]
+{1\over 2}\bar{\psi}\Gamma ^{\mu}[A_{\mu},\psi ]) ,
\label{action}
\eeq
here $\psi$ is a ten dimensional Majorana-Weyl spinor field, and
$A_{\mu}$ and $\psi$ are $N \times N$ Hermitian matrices.
\footnote{
We use the conventions $\{\Gamma^{\mu},\Gamma^{\nu}\}=-2\eta^{\mu\nu}$
where $\eta = $ diag $(-+\ldots +)$.
}
Since this model is formulated in terms of Hermitian matrices,
it has $U(N)$ gauge symmetry.
It has been postulated that the Wilson loops are the vertex operators
for fundamental strings since they are the only gauge invariant observables
\cite{FKKT}.
They appear to play a key role to uncover local symmetries in matrix models.

Non-commutative (NC) gauge theory can be naturally obtained from matrix models
by postulating noncommutative
space-time\cite{CDS}\cite{AIIKKT}\cite{Li}.  The gauge invariant
observables in NC gauge theory,  Wilson lines, have been obtained from
the Wilson loops in matrix models\cite{IIKK}.
The Wilson lines share analogous properties
with string theory such as the UV-IR mixing\cite{MRS}
and universal high energy behavior\cite{Gross}\cite{DK}\cite{Rozali}.
In string theory approach to NC gauge theory\cite{SW},
the Wilson lines couple NC gauge fields to closed string modes
\cite{Liu}$\sim$\cite{Okuyama}.
In bosonic string theory context, we can understand gauge
symmetries of closed string modes from the reparametrization
invariance of the Wilson lines\cite{ADYK}.
We are hence motivated to
investigate analogous problem in superstring context.
However the scope of this paper is limited to massless
modes of superstring, namely the supergravity multiplet.
Although we investigate a matrix model in this paper,
it is straightforward to extend our results to NC gauge theory.

The IIB matrix model possesses
$\cal{N}$ = 2 supersymmetry as
\beqa
\delta^{(1)}\psi &=& \frac{i}{2}
                     [A_{\mu},A_{\nu}]\Gamma^{\mu\nu}\epsilon ^1,\n
\delta^{(1)} A_{\mu} &=& i\bar{\epsilon ^1}\Gamma_{\mu}\psi ,
\label{Ssym1}
\eeqa
and
\beqa
\delta^{(2)}\psi &=& -\epsilon ^2 ,\n
\delta^{(2)} A_{\mu} &=& 0.
\label{Ssym2}
\eeqa

Up to the gauge symmetry
and the equation of motion for $\psi$, we have the following
commutation relations:
\beqa
(\delta^{(1)}_{\epsilon_1}\delta^{(1)}_{\epsilon_2}
    -\delta^{(1)}_{\epsilon_2}\delta^{(1)}_{\epsilon_1})\psi   &=&0 ,\n
(\delta^{(1)}_{\epsilon_1}\delta^{(1)}_{\epsilon_2}
    -\delta^{(1)}_{\epsilon_2}\delta^{(1)}_{\epsilon_1})A_{\mu}&=&0 .
\eeqa
We can also easily check the following commutators:
\beqa
(\delta^{(1)}_{\epsilon}\delta^{(2)}_{\xi}
    -\delta^{(2)}_{\xi}\delta^{(1)}_{\epsilon})\psi   &=&0 ,\n
(\delta^{(1)}_{\epsilon}\delta^{(2)}_{\xi}
    -\delta^{(2)}_{\xi}\delta^{(1)}_{\epsilon})A_{\mu}&=&
                     i\bar{\epsilon}\Gamma_{\mu}\xi ,
\eeqa
and
\beqa
(\delta^{(2)}_{\xi_1}\delta^{(2)}_{\xi_2}
    -\delta^{(2)}_{\xi_2}\delta^{(2)}_{\xi_1})\psi   &=&0 ,\n
(\delta^{(2)}_{\xi_1}\delta^{(2)}_{\xi_2}
    -\delta^{(2)}_{\xi_2}\delta^{(2)}_{\xi_1})A_{\mu}&=&0 .
\eeqa

If we take a linear combination of $\delta^{(1)}$ and $\delta^{(2)}$ as
\beqa
\tilde{\delta}^{(1)}&=&\delta^{(1)}+\delta^{(2)}, \n
\tilde{\delta}^{(2)}&=&{1\over i}(\delta^{(1)}-\delta^{(2)}),
\eeqa
we obtain the $N=2$ supersymmetry algebra,
\beqa
(\tilde{\delta}^{(1)}_{1}\tilde{\delta}^{(1)}_{2}
    -\tilde{\delta}^{(1)}_{2}\tilde{\delta}^{(1)}_{1})
\psi   &=&0 ,\n
(\tilde{\delta}^{(1)}_{1}\tilde{\delta}^{(1)}_{2}
    -\tilde{\delta}^{(1)}_{2}\tilde{\delta}^{(1)}_{1})
A_{\mu}&=& i\bar{\epsilon_1}\Gamma_{\mu}\xi_2
+i\bar{\xi_1}\Gamma_{\mu}\epsilon_2 , \n
(\tilde{\delta}^{(2)}_{1}\tilde{\delta}^{(2)}_{2}
    -\tilde{\delta}^{(2)}_{2}\tilde{\delta}^{(2)}_{1})
\psi   &=&0 ,\n
(\tilde{\delta}^{(2)}_{1}\tilde{\delta}^{(2)}_{2}
    -\tilde{\delta}^{(2)}_{2}\tilde{\delta}^{(2)}_{1})A_{\mu}
&=&i\bar{\epsilon_1}\Gamma_{\mu}\xi_2
+i\bar{\xi_1}\Gamma_{\mu}\epsilon_2 , \n
(\tilde{\delta}^{(1)}_{1}\tilde{\delta}^{(2)}_{2}
    -\tilde{\delta}^{(2)}_{2}\tilde{\delta}^{(1)}_{1})
\psi   &=&0 ,\n
(\tilde{\delta}^{(1)}_{1}\tilde{\delta}^{(2)}_{2}
    -\tilde{\delta}^{(2)}_{2}\tilde{\delta}^{(1)}_{1})
A_{\mu}&=&0 .
\label{Nequal2}
\eeqa
These symmetry considerations force us to interpret
the eigenvalues of $A_{\mu}$ as the space-time coordinates.
We note here that $\epsilon$ and $\xi$ can be
regarded as the complex conjugate to each other.
\footnote{
Such an interpretation is consistent with the fact that
IIB matrix model can be related to Green-Schwarz action
\cite{GS} in the Schild form\cite{Schild}
after the analytic continuation in the fermionic variables\cite{IKKT}.}

The IIB supergravity multiplet consists of a real graviton $h_{\mu\nu}$,
a real fourth rank antisymmetric tensor $A_{\mu\nu\rho\sigma}$,
a complex dilaton $\Phi$, a complex dilatino $\lambda$,
a complex antisymmetric tensor $B_{\mu\nu}$
and a complex gravitino $\eta_{\mu}$.
These fields may be introduced in IIB matrix model as
the sources which couple to the Wilson lines.
They transform in a definite way under local supersymmetry
transformation in IIB supergravity\cite{West}.
The local supersymmetry is realized in the matrix model if
the perturbed theory is invariant under such
a transformation.
We show that it is the case at the linearized level
of the symmetry under a low energy approximation.
This conclusion follows from $\cal{N}$=2 supersymmetry
of IIB matrix model.

In section 2, we briefly recall
current understandings of the Wilson lines
in matrix models and string theory.
In section 3, we investigate the vertex operators
for the supergravity multiplet and the symmetries
of their correlators.
We conclude in section 4 with discussions.

\section{Wilson lines in matrix models and string theory}
\setcounter{equation}{0}
In matrix models, a generic Wilson line is defined along an open contour
$C$ :
\beqa
w(C) & = & Str[\prod_i\hat{O}_i v(C)]  ,\n
v(C) &=&
P exp\{i\int _C d\sigma (k({\sigma})^{\mu}A_{\mu}
 )\} .
\label{Wilsonloop}
\eeqa
Here $k(\sigma)^{\mu}$ denotes the momentum density distributed along
$C$. Since $P$ denotes the path ordering, $Trv(C)$ is obtained
from a familiar Wilson loop operator by assuming the gauge
fields to be constant.
$\hat{O}_i$ denotes matrices such as $\psi$, $[A_{\mu},A_{\nu}]$ ,
$[A_{\mu},\psi]$ or products of them.
The translation invariance requires that $A_{\mu}$ must
appear through the commutators in $\hat{O}_i$.
The symbol $Str$ denotes the symmetric trace which
specifies the ordering of the matrices.
In $Str$, $\hat{O}_i$ should
be treated as a single entity.
Therefore $Str$ amounts to average all possible ways of insertions
of $\hat{O}_i$ into $v(C)$.
Unlike the Wilson loops in gauge theory,
the gauge invariant observables are specified by
generic open contours.

Supergravity type long range interactions emerge
in matrix models after integrating out
off-diagonal components of the matrices.
The Wilson lines appear in such an effective action.
We recall the one loop amplitude of IIB matrix model.
\beqa
W    &=&-{\cal T}r
   \left(\frac{1}{P^2}F_{\mu\nu} \frac{1}{P^2}F_{\nu\lambda}
             \frac{1}{P^2}F_{\lambda\rho}\frac{1}{P^2}F_{\rho\mu} \right) \n
       &~& -2{\cal T}r
     \left(\frac{1}{P^2}F_{\mu\nu} \frac{1}{P^2}F_{\lambda\rho}
             \frac{1}{P^2}F_{\mu\rho}\frac{1}{P^2}F_{\lambda\nu} \right) \n
       &~&+\frac{1}{2}{\cal T}r
   \left(\frac{1}{P^2}F_{\mu\nu} \frac{1}{P^2}F_{\mu\nu}
             \frac{1}{P^2}F_{\lambda\rho}\frac{1}{P^2}F_{\lambda\rho} \right)
\n
      &~&+\frac{1}{4}{\cal T}r
  \left(\frac{1}{P^2}F_{\mu\nu} \frac{1}{P^2} F_{\lambda\rho}
             \frac{1}{P^2}F_{\mu\nu}\frac{1}
{P^2}F_{\lambda\rho} \right)
       +O((F_{\mu\nu})^5).
\label{Wexpansion2}
\eeqa
Here $P_{\mu}$ and $F_{\mu\nu}$ are operators acting on the space of
matrices as
\beqa
        P_{\mu}X & = & [p_{\mu},X] ,\n
        F_{\mu\nu}X & = & \left[ f_{\mu\nu},X \right],
\label{adjointoperator}
\eeqa
where $f_{\mu \nu}=i[p_{\mu},p_{\nu}]$.
In matrix models, long-range interactions may be investigated
by considering the background $p_{\mu}$ to be of the block-diagonal form.
We can then decompose $f_{\mu\nu}=f_{\mu\nu}^i+f_{\mu\nu}^j$.

The long range interactions between well separated and
localized gauge configurations $f_{\mu\nu}^i$ and $f_{\mu\nu}^j$ are
\beqa
&&-{3\over r^8} \left(Str_i[ f_{\mu\nu}f_{\nu\rho}f_{\rho\sigma}f_{\sigma\mu}
-{1\over 4}f_{\mu\nu}f_{\nu\mu}f_{\rho\sigma}f_{\sigma\rho}]
\times Str_j [1]+(i\leftrightarrow j)\right)\n
&&-{12\over r^8} \left(Str_i[ f_{\nu\rho}f_{\rho\sigma}f_{\sigma\mu}
-{1\over 4}f_{\nu\mu}f_{\rho\sigma}f_{\sigma\rho}]
\times Str_j [f_{\mu\nu}]+(i\leftrightarrow j)\right)\n
&&-{12\over r^8}Str_i[ f_{\mu\nu}f_{\nu\rho}]Str_j[f_{\mu\sigma}f_{\sigma\rho}]
+{3\over 2 r^8}Str_i[
f_{\mu\nu}f_{\nu\mu}]Str_j[f_{\rho\sigma}f_{\sigma\rho}]\n
&&+{9\over  r^8}Str_i[ f_{[\mu\nu}f_{\rho\sigma ]}]
Str_j[f_{[\mu\nu}f_{\rho\sigma ]}] ,
\label{blblin}
\eeqa
where $Str_i$ denotes the symmetric trace over the $i$-th sub-matrix space.
$[\mu\nu\rho\sigma ] $ denotes the anti-symmetrization among the indices
with unit weight. We use $(\mu\nu)$ for the symmetrization among the indices
later.
The first line in the above expression can be understood
by the exchange of a complex dilaton field, the second by a complex
second rank antisymmetric tensor field,
the third by a graviton and the last by a fourth rank antisymmetric
tensor respectively.
Here we have not addressed the issue of separating
a dilaton from the trace part of a graviton.
In (\ref{blblin}), we note that the Wilson loops act as
the vertex operators. We can also read-off the bosonic structure of the
relevant vertex operators.
We indeed find the identical structure for the vertex operators
in the following section.
This kind of investigation has been extended with
the inclusion of fermionic backgrounds in
\cite{Taylor}\cite{Schippa}\cite{Kimura}.

In NC gauge theory context, such a phenomena
is called as the UV-IR mixing\cite{MRS}\cite{IKK}.
It can be uniquely traced to the non-planar sector.
In NC gauge theory, the Wilson lines appear in the one loop
effective action through non-planar diagrams.
In fully NC gauge theory, eq.(\ref{Wexpansion2}) can be evaluated as
\beqa
W    &=&V\int {d^Dk\over (2\pi )^D} Tr [ exp(-ik\cdot\hat{x})\n
   &&\times \{
   -\frac{1}{P^2}F_{\mu\nu} \frac{1}{P^2}F_{\nu\lambda}
             \frac{1}{P^2}F_{\lambda\rho}\frac{1}{P^2}F_{\rho\mu} \n
   &&     -2
     \frac{1}{P^2}F_{\mu\nu} \frac{1}{P^2}F_{\lambda\rho}
             \frac{1}{P^2}F_{\mu\rho}\frac{1}{P^2}F_{\lambda\nu}  \n
       &~&+\frac{1}{2}
   \frac{1}{P^2}F_{\mu\nu} \frac{1}{P^2}F_{\mu\nu}
             \frac{1}{P^2}F_{\lambda\rho}\frac{1}{P^2}F_{\lambda\rho}\n
    &&  +\frac{1}{4}
  \frac{1}{P^2}F_{\mu\nu} \frac{1}{P^2} F_{\lambda\rho}
             \frac{1}{P^2}F_{\mu\nu}\frac{1}
{P^2}F_{\lambda\rho}\}  exp(ik\cdot\hat{x}) ]
       +O((F_{\mu\nu})^5),
\label{Wexpansion3}
\eeqa
where we normalize $Tr[\hat{1}]=1$.
We assume
$[\hat{x}^{\mu},\hat{x}^{\nu}]=i\theta^{\mu\nu}$
and the rank of $\theta$ is $D$.
For small external momenta, we can evaluate the above as
\beqa
&&V\theta^{8-D}\int {d^Dq\over (2\pi )^D} \int d^Dy exp(iq\cdot y)\n
&&\times\{-{24\over |y|^8} Tr[exp(-iq\cdot \hat{x}) (
f_{\nu\rho}f_{\rho\sigma}f_{\sigma\mu}
-{1\over 4}f_{\nu\mu}f_{\rho\sigma}f_{\sigma\rho})]
Tr[exp(iq\cdot \hat{x})f_{\mu\nu}]\n
&&-{12\over |y|^8}Tr[exp(-iq\cdot \hat{x}) f_{\mu\nu}f_{\nu\rho}]
Tr[exp(iq\cdot \hat{x})f_{\mu\sigma}f_{\sigma\rho}]\n
&&+{3\over 2 |y|^8}Tr[exp(-iq\cdot \hat{x}) f_{\mu\nu}f_{\nu\mu}]
Tr[exp(iq\cdot \hat{x})f_{\rho\sigma}f_{\sigma\rho}]\n
&&+{9\over  |y|^8}Tr[exp(-iq\cdot \hat{x}) f_{[\mu\nu}f_{\rho\sigma ]}]
Tr[exp(iq\cdot \hat{x})f_{[\mu\nu}f_{\rho\sigma ]}] \} .
\eeqa
Here we can indeed recognize
the Wilson lines which are expanded by NC gauge fields to the
leading nontrivial order.
Various aspects of the Wilson lines in NC field theory have been
studied in\cite{AMNS}\cite{Rey}.

The structure of the Wilson line operators has been also
investigated in superstring theory\cite{OO}.
We can evaluate the one point functions
of the vertex operators for the supergravity multiplet
with gauge fields at the boundary of the disk.
The correlation function which we need to evaluate is
\beq
(z-\bar{z})^2<V^{(-1,-1)}(z)TrPexp\left(i\int dt U^{(0)}(t)\right)> .
\eeq
$U^{(0)}(t)$ is the operator in the $0$-picture for open string
\beq
U^{(0)}(t)
=i\Phi_{\mu}\partial_{\bot}X^{\mu}
-i[\Phi_{\mu},\Phi_{\nu}]\Psi^{\mu}\Psi^{\nu} ,
\eeq
where $\Phi_{\mu}$ are $N\times N$ Hermitian matrices.

Let us consider the vertex operator for a
graviton and an antisymmetric tensor in the $(-1,-1)$-picture
\beq
V^{(-1,-1)}(z)=\delta(\gamma )\delta (\bar{\gamma})
\psi^{\mu}(z)\bar{\psi}^{\nu}(\bar{z})e^{ik\cdot X(z)} .
\eeq
In the appropriate scaling limit, we find
\beqa
&&(z-\bar{z})<\psi^{\mu}(z)\bar{\psi}^{\nu}(\bar{z})e^{ik\cdot X(z)}
TrPexp\left(i\int dt U^{(0)}(t)\right)>\n
&=&Tr[
Pexp\left(i\int_0^1 d\tau k\cdot A\right)\n
&&\times \left({i\over 2\pi}\int_0^1 d\tau_1 [A^{\mu},A^{\nu}]
+{1\over (2\pi )^2\alpha '}\int_0^1 d\tau_1 [A^{\mu},A^{\rho}]
\int_0^1 d\tau_2[A_{\rho},A^{\nu}] \right)] ,
\eeqa
where $A^{\mu}=2\pi\alpha '\Phi^{\mu}$.
In this way we can determine the bosonic part of the
vertex operators for $B_{\mu\nu}$ and $h_{\mu\nu}$.
The results are consistent with
the one loop effective action of NC gauge theory.

Let us consider the simplest gauge invariant operator
which corresponds to a straight Wilson line operator
in NC gauge theory:
\beq
Tr[exp (ik\cdot A)] .
\eeq
In order to understand the SUSY multiplet to which
it belongs,
we consider the SUSY transformation eq.(\ref{Ssym1}).
We observe that
the SUSY transformation $\delta^{(1)}$ whose generator satisfies
$k_{\mu}\Gamma^{\mu}\epsilon=0$
commutes with the Wilson line. If $k$ is a null vector, the straight Wilson
line
is a BPS state since it preserves the half of SUSY.
Let us choose the Lorentz frame such that $k_{\mu}\Gamma^{\mu}
=k^{+}\Gamma^{-}$. The broken SUSY generators satisfy
$\Gamma^{+}\lambda=0$. The massless SUSY multiplets are generated by such
broken SUSY generators $\{\lambda^a\}$ from the null Wilson line.
They can be represented by a superfield $\Psi$
which is a polynomial in $\lambda^a$ up to the eighth order.
\beqa
\Psi&=&A+\psi^{a}\lambda^{a}+
{1\over 2}A^{ab}\lambda^a\lambda^b\n
&&-{1\over 3!}\psi^{abc}\lambda^a\lambda^b\lambda^c+
{1\over 4!}A^{abcd}\lambda^a\lambda^b\lambda^c\lambda^d\n
&&+{1\over 3!\cdot 5!}\psi^{abc*}\epsilon^{a\ldots h}\lambda^d\cdots\lambda^h-
{1\over 2\cdot 6!}A^{ab*}\epsilon^{a\ldots h}\lambda^c\cdots\lambda^h\n
&&+{1\over 7!}\psi^{a*}\epsilon^{a\ldots h}\lambda^b\cdots\lambda^h+
{1\over 8!}A^*\epsilon^{a\ldots h}\lambda^a\cdots\lambda^h .
\eeqa
It is in fact known that these $2^8$ fields form the IIB
supergravity multiplet \cite{BGS} .
The light-cone gauge formulation of Wilson loops can be
extended to massive states in IIB matrix model\cite{Hamada}.

The merit of this light-cone type argument is
its robustness against possible quantum
corrections due to the BPS nature of the multiplet.
The drawback is that it is oblivious to gauge symmetries.
We may draw an analogy with $U(1)$ gauge theory here.
The on-shell photon has two transverse degrees of freedom.
Its covariant off-shell extension is described by $U(1)$ gauge theory.
It is most likely that the covariant off-shell extension of the straight
Wilson lines and their descendants is described by supergravity.
In the following section, we identify such expected local symmetries.

\section{$N=2$ SUSY and the vertex operators}
\setcounter{equation}{0}
In this section, we construct the vertex operators in IIB matrix model
which couple to the supergravity multiplet.
The relevant operators can be constructed
through the Wilson lines.
We start with the simplest operator which
corresponds to a straight Wilson line in NC gauge theory:
\beq
Tr[exp(ik\cdot A)]\Phi (k) .
\label{dilaton}
\eeq
In bosonic string theory, $\Phi (k)$ has been interpreted as a tachyon field
with momentum $k$.
A natural question is how to interpret this operator in superstring.
In this section we argue that $\Phi$ can be interpreted as a
dilaton field in the IIB supergravity multiplet.

Our strategy is to consider the generating functional of the
Wilson line correlators:
\beq
e^{W(\Phi_i)}=<e^{ V_i\Phi_i}> ,
\label{linear}
\eeq
where the average is taken with respect to the IIB matrix model action
given in eq.(\ref{action}). $\Phi_i$ and $V_i$ denote the fields and dual
vertex operators descended from $\Phi (k)$ and $Tr[exp(ik\cdot A)]$
by $\cal{N}$=2 SUSY  .
We will shortly find that they form the IIB supergravity
multiplet as expected.
By considering the change of variables which
is identical to SUSY transformation in (\ref{Ssym1}) and
(\ref{Ssym2}) ,
we can derive the Ward identities in a standard way:
\beq
<e^{V_i\Phi_i}>
=<e^{V_i(A+\delta A,\psi+\delta\psi )\Phi_i}> .
\eeq

Our task is to construct the vertex operators $\{V_i\}$ in order to
satisfy the following relation with the judicious choice of $\delta\Phi_i$
\footnote
{This strategy has been employed in several other contexts.
See \cite{Nicolai} for a Matrix theory application. }
\beq
<e^{V_i(A+\delta A,\psi+\delta\psi )\Phi_i}>
=<e^{V_i(\Phi_i +\tilde\delta\Phi_i )}> .
\label{reqmnt}
\eeq
If it is successful, we can derive the symmetry of the generating
functional for the correlators of Wilson lines:
\beq
e^{W(\Phi_i)}=e^{W(\Phi_i +\tilde\delta\Phi_i)} .
\label{sugrtr}
\eeq
Since our scope is limited to the supergravity multiplet
in this paper, we carry out this program
under a low energy approximation.
We find that the symmetry of $W(\Phi_i)$ coincides with the local
SUSY transformation of IIB supergravity at the linearized level\cite{West}.
We denote it as SUGRA transformation in what follows.

With these motivations,
let us apply SUSY transformation to eq.(\ref{dilaton}):
\beq
\delta Tr [exp(ik\cdot A)]\Phi (k) =
Tr [exp(ik\cdot A)\bar{\psi}]k\cdot \Gamma\epsilon ^1 \Phi (k) .
\label{Ssytr1}
\eeq
In order to satisfy eq.(\ref{reqmnt}),
we are led to add a new operator to
eq.(\ref{dilaton}) which contains the dilatino field
$\lambda$:
\beqa
&&-Tr [exp(ik\cdot A)\bar{\psi}]\lambda ,\n
&&\tilde{\delta}\lambda = -k\cdot \Gamma\epsilon ^1 \Phi (k) .
\label{dltnvt}
\eeqa
Through eq.(\ref{sugrtr}), we have identified
SUGRA transformation of the dilatino field as (\ref{dltnvt}).

We in turn need to consider SUSY transformation of the newly
introduced vertex operator:
\beqa
&&-\delta Tr [exp(ik\cdot A)\bar{\psi}]\lambda\n
&=&Tr [exp(ik\cdot A)]\bar{\epsilon ^2}\lambda\n
&&+Tr [exp(ik\cdot A)\frac{i}{2}[A_{\mu},A_{\nu}]]\bar{\epsilon ^1}
\Gamma^{\mu\nu}\lambda
-Str [exp(ik\cdot A)\bar{\psi}k\cdot \Gamma\epsilon ^1 \bar{\psi}\lambda ] .
\label{dilntr}
\eeqa
The symmetric trace, $Str$ implies the following operation
\beqa
&&Str [exp(ik\cdot A)\psi_{\alpha}\psi_{\beta}]\n
&=&\int_0^1 d\sigma \int_{\sigma}^1 d\sigma '
Tr [exp(i\sigma k\cdot A))\psi_{\alpha}exp(i(\sigma '-\sigma ) k\cdot A))
\psi_{\beta}exp(i(1-\sigma ' )k\cdot A))]\n
&&- (\alpha  \leftrightarrow \beta ) ,
\label{symtr}
\eeqa
where the $-$ sign is due to the fermionic nature of $\psi$.
Starting from the simply traced object in eq.(\ref{dilaton}),
we only obtain symmetric traced objects.
We can further recombine
fermionic variables through
the following Fierz identity
\beqa
\bar{\phi}_{\beta}\psi_{\alpha}
&=&-{1\over 16} \Gamma^{\mu}_{\alpha\beta}\bar{\phi}\Gamma_{\mu}\psi\n
&&+{1\over 16\cdot 3!}
\Gamma^{\mu\nu\rho}_{\alpha\beta}\bar{\phi}\Gamma_{\mu\nu\rho}\psi\n
&&-{1\over 16\cdot 5!}
\Gamma^{\mu\nu\rho\sigma\tau}_{\alpha\beta}\bar{\phi}
\Gamma_{\mu\nu\rho\sigma\tau}\psi ,
\eeqa
where $\psi$ and $\phi$ are Majorana-Weyl spinors in ten dimensions.

The first term in eq.(\ref{dilntr}) can be interpreted
as the SUGRA transformation of a dilaton:
\beq
\tilde\delta \Phi = \bar{\epsilon ^2}\lambda .
\eeq
The remaining terms lead us
to introduce the coupling with the antisymmetric tensor field
$B^{\mu\nu}$ in addition
\beqa
&&Tr exp(ik\cdot A)\frac{i}{2}[A_{\mu},A_{\nu}]B^{\mu\nu}
-{i\over 48}Str exp(ik\cdot A)\bar{\psi}\Gamma_{\mu\nu\rho}\psi
H^{\mu\nu\rho},\n
&&\tilde\delta B^{\mu\nu} = \bar{\epsilon ^1}
\Gamma^{\mu\nu}\lambda ,\n
&&H_{\mu\nu\rho}\equiv \partial_{\rho}B_{\mu\nu}
+\partial_{\mu}B_{\nu\rho}+\partial_{\nu}B_{\rho\mu} .
\label{bmunu}
\eeqa
Here only the anti-symmetric part in the fermion bilinear
term contributes due to the definition of $Str$ as in eq.(\ref{symtr}).
The application of $\partial_{\mu}$ should be understood as the
multiplication by $ik_{\mu}$ in this paper.
We have also assumed that $\Gamma^{\mu}\partial_{\mu}\lambda=0$.
We note that the vertex operator in eq.(\ref{bmunu})
can be shown to be invariant
under the following gauge transformation of $B_{\mu\nu}$ field
\beq
\tilde\delta B_{\mu\nu} = \partial_{[\mu}\Lambda_{\nu ]} .
\eeq

The SUSY transformation of the vertex operator in
eq.(\ref{bmunu}) gives rise to
\beqa
&&{i\over 24}Tr exp(ik\cdot A)\bar{\epsilon^2}\Gamma_{\mu\nu\rho}\psi
H^{\mu\nu\rho} \n
&&+{1\over 48} Str exp(ik\cdot
A)\bar{\psi}[A_{\alpha},A_{\beta}]\Gamma^{\alpha}
(\Gamma^{\beta\nu\rho\lambda}H_{\nu\rho\lambda}
+9\Gamma^{\rho\lambda}H_{\beta\rho\lambda})\epsilon^1\n
&&-{i\over 48}Str exp(ik\cdot A)\bar{\psi}k\cdot \Gamma \epsilon^1
\bar{\psi}\Gamma_{\mu\nu\rho}\psi
H^{\mu\nu\rho} .
\label{bmntr}
\eeqa
The first term of eq.(\ref{bmntr}) can be obtained from
eq.(\ref{dltnvt}) by
postulating the following SUGRA transformation for
dilatino field:
\beq
\tilde\delta \lambda
=-{i\over 24}\Gamma_{\mu\nu\rho}\epsilon^2H^{\mu\nu\rho} .
\eeq
The second term necessitates us
to introduce the following vertex operator which couples to
gravitino field $\eta_{\mu}$:
\beqa
&&Str exp(ik\cdot A)\bar{\psi}[A_{\alpha},A_{\beta}]
\Gamma^{\alpha}\eta^{\beta},\n
&&\tilde\delta \eta^{\beta}={1\over 48}
(\Gamma^{\beta\nu\rho\lambda}H_{\nu\rho\lambda}
+9\Gamma_{\rho\lambda}H^{\beta\rho\lambda})\epsilon^1 .
\label{grtnvt}
\eeqa
We can show that this coupling is invariant under local SUSY transformation
$\tilde\delta \eta_{\mu} = -ik_{\mu}\epsilon^2$ by using the
equations of motion for $\psi$.

We are left with the third term in (\ref{bmntr}).
It is $O(k)$ in comparison to the rest
and hence we may argue that it is irrelevant in the low energy limit.
It might be explained by introducing a massive fermionic mode
${\Psi}^{[\mu\nu]}$ which is expected at the first excited
level in superstring as follows
\beqa
&&{1\over 16}Str exp(ik\cdot A)
\bar{\psi}\Gamma_{\mu\nu\rho}\psi\bar{\psi}
\partial^{[\rho}{\Psi}^{\mu\nu ]},\n
&&\tilde {\delta} {\Psi}^{[\mu\nu
]} = -ik\cdot\Gamma B^{\mu\nu}\epsilon^1 .
\label{qdrcpl}
\eeqa
Although it is a very interesting problem to understand
the vertex operators for massive modes, it is beyond the
scope of this paper. We are content to postpone this question
by arguing that it is irrelevant in the low energy approximation.

The SUSY transformation of the gravitino coupling in eq.(\ref{grtnvt})
gives rise to:
\beqa
&&
-Tr exp(ik\cdot A)
[A_{\alpha},A_{\beta}]\bar{\epsilon^2}\Gamma^{\alpha}\eta^{\beta}\n
&&+{i}Strexp(ik\cdot A)
[A_{\alpha},A_{\beta}][A^{\beta},A_{\mu}]
\bar{\epsilon^1}\Gamma^{\alpha}\eta^{\mu}\n
&&-{i\over 2}Str exp(ik\cdot A)
[A_{\alpha},A_{\beta}][A_{\gamma},A_{\mu}]
\bar{\epsilon^1}\Gamma^{\alpha\beta\gamma}\eta^{\mu}\n
&&+{i\over 2}Strexp(ik\cdot A)\bar{\psi}\Gamma_{\alpha}
[\psi,A_{\mu}]\bar{\epsilon^1}\Gamma^{\alpha}\eta^{\mu}\n
&&+{1\over 48}Strexp(ik\cdot A)\bar{\psi}\Gamma_{\alpha\beta\gamma}
\psi[A_{\mu},k\cdot A]\bar{\epsilon^1}\Gamma^{\alpha\beta\gamma}\eta^{\mu}\n
&&+Strexp(ik\cdot A) \bar{\epsilon^1}\Gamma_{\rho}\psi
[A_{\alpha},A_{\beta}]\bar{\psi}(k^{\beta}\Gamma^{\alpha}\eta^{\rho}
-k^{\rho}\Gamma^{\alpha}\eta^{\beta}) .
\label{strgrv}
\eeqa
The first term in eq.(\ref{strgrv})
can be obtained from eq.(\ref{bmunu}) by assuming the
following transformation for $B_{\mu\nu}$ field
\beq
\tilde\delta B_{\mu\nu}=
2i\bar{\epsilon^2}\Gamma_{[\mu}\eta_{\nu ]} .
\eeq

The terms in the last two lines in eq.(\ref{strgrv}) are $O(k)$.
We also note that the antisymmetric part of the
following term is also $O(k)$ since
\beqa
&&{i\over 4}Strexp(ik\cdot A)\bar{\psi}
(\Gamma_{\alpha}[\psi,A_{\mu}]-\Gamma_{\mu}[\psi,A_{\alpha}])
\bar{\epsilon^1}\Gamma^{\alpha}\eta^{\mu}\n
&&={i\over 8}Strexp(ik\cdot A)\bar{\psi}
(\Gamma^{\beta}\Gamma_{\alpha\mu}-\Gamma_{\alpha\mu}\Gamma^{\beta} )
[\psi,A_{\beta}]\bar{\epsilon^1}\Gamma^{\alpha}\eta^{\mu}\n
&&={i\over 8}Strexp(ik\cdot A)\bar{\psi}\Gamma^{\beta\alpha\mu}\psi
[A_{\beta},ik\cdot A]\bar{\epsilon^1}\Gamma_{\alpha}\eta_{\mu} ,
\eeqa
where we have used the equation of motion.

In order to account for eq.(\ref{strgrv}) in terms of SUGRA
transformations,
we need to introduce the Wilson lines which couple to
a graviton $h_{\alpha\mu}$ and
a four-th rank antisymmetric field $A_{\mu\nu\rho\sigma}$:
\beqa
&&Str exp(ik\cdot A)
([A^{\alpha},A^{\beta}][A^{\mu},A_{\beta}]
+{1\over 2}\bar{\psi}\Gamma^{(\alpha}[A^{\mu )},\psi])
h_{\alpha\mu}\n
&&+ {1\over 2}Str exp(ik\cdot A) \bar{\psi}\Gamma^{\rho\beta(\alpha}\psi
[A^{\mu )},A_{\beta}]\partial_{\rho}h_{\alpha\mu}\n
&&-{i\over 2}Str exp(ik\cdot A)
[A^{\alpha},A^{\beta}][A^{\mu},A^{\nu}]A_{\alpha\beta\mu\nu},\n
&&\tilde\delta h_{\alpha\mu} =-i\bar{\epsilon^1}
\Gamma_{(\alpha}\eta_{\mu ) },\n
&&\tilde\delta A_{\alpha\beta\mu\nu}=\bar{\epsilon^1}
\Gamma_{[\alpha\beta\mu}\eta_{\nu ]} .
\label{gravtx}
\eeqa
We can show that the couplings in eq.(\ref{gravtx}) are invariant under
the following gauge transformations after using the equation of motion
\beqa
\tilde\delta h_{\alpha\mu}&=&\partial_{\mu}\xi_{\alpha }
+\partial_{\alpha}\xi_{\mu },\n
\tilde\delta A_{\mu\nu\rho\sigma} &=&
\partial_{[\mu}\Lambda_{\nu\rho\sigma ]} .
\eeqa
We have fixed
the $O(k)$ term in eq.(\ref{gravtx}) in order to satisfy
the invariance with respect to the general coordinate
transformation.
After this procedure, there remains the following
discrepancy between
SUGRA transformation of (\ref{gravtx}) and (\ref{strgrv})
\beqa
&&Strexp(ik\cdot A)\{{1\over 48}\bar{\psi}\Gamma_{\alpha\beta\gamma}
\psi[A_{\mu},A^{\rho}]
\bar{\epsilon^1}k_{\rho}\Gamma^{\alpha\beta\gamma}\eta^{\mu}\n
&&+
\bar{\epsilon^1}k_{\beta}\Gamma_{\rho}\psi
[A_{\alpha},A^{\beta}]\bar{\psi}\Gamma^{\alpha}\eta^{\rho}
-\bar{\epsilon^1}k_{\rho}\Gamma^{\rho}\psi
[A_{\alpha},A_{\beta}]\bar{\psi}\Gamma^{\alpha}\eta^{\beta}\n
&&
-{1\over 8}\bar{\psi}\Gamma^{\beta\alpha\mu}\psi
[A_{\beta},A^{\rho}]\bar{\epsilon^1}k_{\rho}\Gamma_{\alpha}\eta_{\mu}
-{1\over 2}\bar{\psi}\Gamma^{\rho\beta(\alpha}\psi
[A^{\mu )},A_{\beta}]\bar{\epsilon^1}k_{\rho}\Gamma_{\alpha}\eta_{\mu}\} .
\label{apdx2}
\eeqa
Although it is necessary to interpret these terms,
it will be left to future investigations since
we can ignore $O(k)$ terms under the low energy approximation.

Although $h_{\mu\alpha}$ and $A_{\mu\nu\rho\sigma}$ are real fields,
$\cal{N}$=2 SUSY requires the introduction of the
complex $B_{\mu\nu}$, dilaton and dilatino fields.
We can in principle determine the structure of the Wilson lines dual
to the complex conjugate of them by repeating this process.
The SUSY transformation of eq.(\ref{gravtx}) is investigated in
Appendix. Although we have shown that the result is $O(k)$,
our physical interpretation of it is
still incomplete.

However we believe that the structure of the anti-gravitino
vertex operator can be fixed by the requirement of gauge invariance.
Here we make use of the fact that
we can construct another conserved fermionic current
in addition to that in (\ref{grtnvt}).
We propose to couple
$\bar{\eta}^c_{\mu}$ to such a current as follows:
\beqa
&&-{i\over 2}Str exp(ik\cdot A)\bar{\eta}^c_{\mu}
[A^{\mu},A_{\nu}][A_{\rho},A_{\sigma}]
\Gamma^{\rho\sigma}\Gamma^{\nu}\psi\n
&=&-iStr exp(ik\cdot A)\bar{\eta}^c_{\mu}
[A^{\mu},A_{\nu}][A^{\nu},A_{\sigma}]
\Gamma^{\sigma}\psi\n
&&-{i\over 2}Str exp(ik\cdot A)\bar{\eta}^c_{\mu}
[A^{\mu},A_{\nu}][A_{\rho},A_{\sigma}]
\Gamma^{\nu\rho\sigma}\psi .
\label{antgrv}
\eeqa
We can show the gauge invariance of the coupling
corresponding to local SUSY
$\delta \bar{\eta}^c_{\mu}=ik_{\mu}\bar{\epsilon^1}$ by
using the equation of motion and neglecting
cubic terms in $\psi$.

The SUSY transformation of eq.(\ref{antgrv}) leads to
\beqa
&&iStr exp(ik\cdot A)\bar{\eta}^c_{\mu}
[A^{\mu},A_{\nu}][A^{\nu},A_{\sigma}]
\Gamma^{\sigma}\epsilon^2\n
&&+{i\over 2}Str exp(ik\cdot A)\bar{\eta}^c_{\mu}
[A^{\mu},A_{\nu}][A_{\rho},A_{\sigma}]
\Gamma^{\nu\rho\sigma}\epsilon^2\n
&&-2
Str exp(ik\cdot A)[A^{\mu},A_{\nu}][A^{\nu},A_{\sigma}][A^{\sigma},A_{\alpha}]
\bar{\eta}^c_{\mu}\Gamma^{\alpha}\epsilon^1\n
&&+{1\over2}
Str exp(ik\cdot A)[A^{\mu},A_{\nu}][A_{\rho},A_{\sigma}][A^{\sigma},A^{\rho}]
\bar{\eta}^c_{\mu}\Gamma^{\nu}\epsilon^1 ,
\eeqa
where we have neglected fermionic terms.
It in turn implies
\beqa
&&\tilde\delta h_{\alpha\mu} =
-i\bar{\eta}^c_{(\mu}\Gamma_{\alpha )}\epsilon ^2,\n
&&\tilde\delta A_{\alpha\beta\mu\nu}=-\bar{\eta}^c_{[\alpha}
\Gamma_{\beta\mu\nu ]}\epsilon ^2 .
\eeqa
We also need to introduce the coupling to ${B}^c_{\mu\nu}$ field:
\beqa
&&-i{B}^c_{\mu\alpha}Str exp(ik\cdot A)(
[A^{\mu},A_{\nu}][A^{\nu},A_{\sigma}][A^{\sigma},A^{\alpha}]
-{1\over4}
[A^{\mu},A^{\alpha}][A_{\rho},A_{\sigma}][A^{\sigma},A^{\rho}]),
\n
&&\tilde\delta {B}^c_{\mu\nu}= -2i\bar{\eta}^c_{[\mu}\Gamma_{\nu ]}
\epsilon^1 .
\eeqa
This coupling can be shown to be gauge invariant after using the
equation of motion.

Since we have neglected fermionic terms in the $B^c_{\mu\nu}$
vertex operator, we can no longer
determine the anti-dilatino vertex operator by SUSY
transformations from it.
In order to proceed further, we now resort to use consistency
arguments.
Firstly we expect to obtain the SUGRA transformations
which are complex conjugate to the known type.
Secondly we also expect to obtain the vertex operators which are consistent
with the one loop effective action in section 2.
We may postulate
the anti-dilatino coupling as follows
since it satisfies these consistency requirements,
\beqa
&&-i\bar{\lambda}^c\Gamma^{\mu\nu}Str exp(ik\cdot A)\psi(
[A_{\mu},A_{\rho}][A^{\rho},A_{\sigma}][A^{\sigma},A_{\nu}]
-{1\over4}
[A_{\mu},A_{\nu}][A_{\rho},A_{\sigma}][A^{\sigma},A^{\rho}])
\n
&&-{i\over 30}\bar{\lambda}^c\Gamma^{\mu\nu\rho\sigma\tau\lambda}
Str exp(ik\cdot A)\psi
[A_{\mu},A_{\nu}][A_{\rho},A_{\sigma}][A_{\tau},A_{\lambda }] .
\eeqa

With this ansatz,
the SUSY transformation of the anti-dilatino coupling results in
\beqa
&&i\bar{\lambda}^c\Gamma^{\mu\nu}\epsilon^2
Str exp(ik\cdot A) \n
&&\times ([A_{\mu},A_{\rho}][A^{\rho},A_{\sigma}][A^{\sigma},A_{\nu}]
-{1\over4}
[A_{\mu},A_{\nu}][A_{\rho},A_{\sigma}][A^{\sigma},A^{\rho}]))\n
&&+{i\over 30}\bar{\lambda}^c\Gamma^{\mu\nu\rho\sigma\tau\lambda}
\epsilon^2 Str exp(ik\cdot A)
[A_{\mu},A_{\nu}][A_{\rho},A_{\sigma}][A_{\tau},A_{\lambda}]\n
&&
+\bar{\lambda}^c\epsilon^1 Str exp(ik\cdot A)\n
&&\times (
[A_{\mu},A_{\nu}][A^{\nu},A_{\sigma}][A^{\sigma},A_{\alpha}]
[A^{\alpha},A^{\mu}]
-{1\over4}
[A_{\mu},A_{\alpha}][A^{\alpha},A^{\mu}]
[A_{\rho},A_{\sigma}][A^{\sigma},A^{\rho}])\n
&&
+{1\over 60}\bar{\lambda}^c\Gamma^{\mu\nu\rho\sigma\tau\lambda
\alpha\beta}\epsilon^1
Str exp(ik\cdot A)[A_{\alpha},A_{\beta}]
[A_{\mu},A_{\nu}][A_{\rho},A_{\sigma}][A_{\tau},A_{\lambda}] .
\eeqa
Let us focus on $\epsilon^2$ dependent part first.
The second rank antisymmetric tensor part can be explained by
the following SUGRA transformation
\beq
\tilde\delta {B}^c_{\mu\nu}=-\bar{\lambda}^c\Gamma_{\mu\nu}\epsilon^2 .
\eeq
The sixth rank antisymmetric tensor part presumably
requires the introduction of the sixth rank tensor
field. We may interpret it as a massive mode since
we expect such a mode at the first excited level in superstring.
It possesses the required gauge symmetry to remove the negative
norm states.

Let us move on to $\epsilon^1$ dependent part.
The scalar part requires us to include the anti-dilaton coupling as
\beqa
&&-{\Phi}^cStr exp(ik\cdot A)(
[A_{\mu},A_{\nu}][A^{\nu},A_{\sigma}][A^{\sigma},A_{\alpha}]
[A^{\alpha},A^{\mu}]
-{1\over4}
[A_{\mu},A_{\alpha}][A^{\alpha},A^{\mu}]
[A_{\rho},A_{\sigma}][A^{\sigma},A^{\rho}]),
\n
&&\tilde\delta {\Phi}^c=-\bar{\lambda}^c\epsilon^1 .
\eeqa
The eighth rank antisymmetric tensor part again
may be dealt with by introducing a corresponding massive mode.
It possesses the required gauge symmetry.
Since we have exhausted the entire super gravity multiplet,
this concludes our heuristic arguments to
construct the vertex operators dual
to the IIB supergravity multiplet.

Here we summarize the SUGRA transformations of the source fields
which have been identified in this section.
They are the symmetry of the generating functional
of the Wilson line correlators, $W(\Phi_i)$.
Admittedly the evidence becomes weaker for the lower members of
the list since we can only offer consistency arguments for them.
\beqa
&&\tilde\delta \Phi =\bar{\epsilon^2}\lambda ,\n
&&\tilde\delta\lambda = i\Gamma^{\mu}\epsilon^1 \partial_{\mu} \Phi
-{i\over 24}\Gamma_{\mu\nu\rho}\epsilon^2H^{\mu\nu\rho},\n
&&\tilde\delta B_{\mu\nu}= \bar{\epsilon^1}\Gamma_{\mu\nu}\lambda
+2i\bar{\epsilon^2}\Gamma_{[\mu}\eta_{\nu]},\n
&&\tilde\delta \eta_{\mu}= {1\over 48}
(\Gamma_{\mu\nu\rho\lambda}H^{\nu\rho\lambda}
+9\Gamma^{\rho\lambda}H_{\mu\rho\lambda})\epsilon^1
-\partial_{\mu} \epsilon^2
-{1\over 2}\Gamma^{\nu\rho}\partial_{\nu}h_{\mu\rho}\epsilon^2
+{i\over 4\cdot 5!} \Gamma^{\rho_1\cdots\rho_5} \Gamma_{\mu}{\epsilon^2}
F_{\rho_1\cdots\rho_5},\n
&&\tilde\delta h_{\alpha\mu} =-i\bar{\epsilon^1}\Gamma_{(\alpha}\eta_{\mu ) }
+i\bar{\epsilon ^2}\Gamma_{(\alpha}{\eta}^c_{\mu )},\n
&&\tilde\delta A_{\alpha\beta\mu\nu}=\bar{\epsilon^1}
\Gamma_{[\alpha\beta\mu}\eta_{\nu]}
+\bar{\epsilon^2}\Gamma_{[\alpha\beta\mu}{\eta}^c_{\nu ]},\n
&&\tilde\delta \bar{\eta}^c_{\mu}=\partial_{\mu}\bar{\epsilon^1}
-{1\over 2}\bar{\epsilon^1}\Gamma^{\nu\rho}\partial_{\nu}h_{\mu\rho}
+{i\over 4\cdot 5!}\bar{\epsilon^1}\Gamma_{\mu}
\Gamma^{\rho_1\cdots\rho_5} F_{\rho_1\cdots\rho_5},\n
&&\tilde\delta B^c_{\mu\nu}=
-\bar{\epsilon^2}\Gamma_{\mu\nu}\lambda^c
-2i\bar{\epsilon^1}\Gamma_{[\mu }\eta^c_{\nu ]},\n
&&\tilde\delta \Phi^c = -\bar{\epsilon^1}\lambda^c .
\label{linetr}
\eeqa
These transformations agree with the linearized
local SUSY transformations of IIB supergravity
under the identification that $\epsilon^1=\epsilon^*$ and
$\epsilon^2=\epsilon$ \cite{West}.
\footnote
{We remark that our metric $\eta^{\mu\nu}$ is of the opposite sign
of that in \cite{West}.
We also recall that $\bar{\epsilon}\equiv\epsilon\Gamma^0$
in our conventions.
After rescaling the gravitino field $\eta_{\mu}
\rightarrow 2\eta_{\mu}$,
our SUGRA transformations coincide with their $\kappa =2$ case.}

The commutator of two local supersymmetry transformations
supposed to give all six types of local symmetry transformations:
\beq
[\tilde\delta  (\epsilon_1 ),\tilde\delta  (\epsilon_2 )]
=i\tilde\delta (x)+i\tilde\delta (l)+i\tilde\delta (\epsilon )
+i\tilde\delta (\Lambda_{\cdot}) +i\tilde\delta (\Lambda_{\cdot\cdot\cdot})
+i\tilde\delta (\Sigma ) ,
\label{comssy}
\eeq
where we have listed the general coordinate
$\tilde\delta (x)$, local Lorentz $\tilde\delta (l)$,
local supersymmetry $\tilde\delta (\epsilon )$,
gauge transformations for the second $\tilde\delta (\Lambda_{\cdot})$
and the fourth rank tensor $\tilde\delta (\Lambda_{\cdot\cdot\cdot})$
and $U(1)$ transformation $\tilde\delta (\Sigma )$
respectively.
We can check eq.(\ref{comssy}) at the linearized level with the
transformations
in eq.(\ref{linetr}) as follows:

general-coordinate transformation with
$x^{\mu}=\bar{\epsilon}_2^*\Gamma^{\mu}\epsilon_1
+\bar{\epsilon}_2\Gamma^{\mu}\epsilon_1^* $;

gauge transformation for $B_{\cdot}$  with
$\Lambda_{\mu}=2\bar{\epsilon_2}\Gamma_{\mu}\epsilon_1$;

gauge transformation for $A_{\cdot\cdot\cdot}$  with
$\Lambda_{\alpha\beta\gamma}=({1/ 4i})\bar{\epsilon_2}^*
\Gamma_{\alpha\beta\gamma}\epsilon_1
-({1/ 4i})\bar{\epsilon_2}
\Gamma_{\alpha\beta\gamma}\epsilon_1^*$.

\section{Conclusions and discussions}
We have investigated the Wilson lines which couple
to the supergravity multiplet in IIB matrix model.
These fields may be introduced as
the sources which couple to the Wilson lines.
They transform in a definite way
under the local supersymmetry transformation.
The local supersymmetry is realized in matrix models if
the perturbed theory is invariant under such
a transformation.
We have shown that it is the case at the linearized level
of the symmetry for a class of massless modes.
This conclusion follows from $\cal{N}$=2 supersymmetry
of IIB matrix model.

In matrix models, there is no distinction between
NS-NS and R-R fields. For example the graviton $h_{\mu\nu}$
is a NS-NS field while the fourth rank antisymmetric tensor
$A_{\mu\nu\rho\sigma}$ is a R-R field.
They both appear in eq.(\ref{gravtx}).
In the case of the complex fields, there are two
different vertex operators for each complex field.
In the case of $B_{\mu\nu}$ field, a T-duality
argument suggests that the  R-R field
couples to the Wilson line which contains the single power of
$[A_{\mu},A_{\nu}]$\cite{Myers}.
On the other hand it also couples to the NS-NS field as it was recalled in
section 2. So we may interpret that $B_{\mu\nu}$ which couples to
$[A_{\mu},A_{\nu}]$ is the linear combination of NS-NS and R-R field  as
$B_{NS}+iB_R$. With such an interpretation,
the one loop effective action eq.(\ref{blblin})
implies that $B^c_{\mu\nu}$ which couples to the third power of
$[A_{\mu},A_{\nu}]$ is of $B_{NS}-iB_R$ type.
Presumably the situation is analogous for other complex fields.

Since we have relied on
the low energy approximation in this paper, our next goal is
to understand the structure of the Wilson lines for massive
modes of superstring.
It looks likely that we have scratched
the vertex operators for massive modes
already in this paper. Therefore
we may be able to understand them by
further extending this work.
Another possibly related problem is
to understand the fully nonlinear structure
of the local SUSY transformations beyond the linearized approximation.

\setcounter{equation}{0}

\begin{center} \begin{large}
Acknowledgments
\end{large} \end{center}
I would like to thank N. Ishibashi, S. Iso, H. Kawai ,S.-J. Rey and
Y. Shibusa for discussions.
This work is supported in part by the Grant-in-Aid for Scientific
Research from the Ministry of Education, Science and Culture of Japan.
A part of this work was carried out while I participated in
the workshop `Mathematical Aspects of String Theory' at
ESI in Vienna.

\appendix
\section{Appendix}

In this section, we investigate SUSY transformation
of the Wilson lines which couple to a graviton
and a four-th rank antisymmetric tensor.
We show that the $O(1)$ terms cancel and the result
is $O(k)$. We then discuss whether we can understand
the result in terms of SUGRA transformations of
known or unknown vertex operators.
Although our physical understanding of it is still incomplete,
we believe it
is worthwhile to report our results for future investigations.

The SUSY transformation of eq.(\ref{gravtx}) gives rise to:
\beqa
&&Str exp(ik\cdot A)
\{i[A^{\alpha},\bar{\epsilon^1}\Gamma^{\beta}\psi][A^{\mu},A_{\beta}]
+{i\over 2}\bar{\epsilon^1}\Gamma^{\alpha\beta\gamma}\psi[A^{\mu},
[A_{\beta},A_{\gamma}]]\n
&&+2\bar{\epsilon^1}\Gamma^{\alpha}\psi[k\cdot A,A^{\beta}][A^{\mu},A_{\beta}]
-\bar{\epsilon^1}k\cdot\Gamma\psi[A^{\alpha}, A^{\beta}][A^{\mu},A_{\beta}]\n
&&-{1\over 4}\bar{\epsilon^1}\Gamma^{\alpha\beta\gamma}\psi
[A^{\mu},k\cdot A][A_{\beta},A_{\gamma}]
-{1\over 2}\bar{\epsilon^1}\Gamma^{\beta}\psi
[A^{\mu},k\cdot A][A_{\beta},A^{\alpha}]\}h_{\alpha\mu}\n
&&-Str exp(ik\cdot A)
\{{i\over 2}\bar{\epsilon^2}\Gamma^{\alpha}\psi[k\cdot A,A^{\mu}]h_{\alpha\mu}
+\bar{\epsilon^2}\Gamma^{\rho\beta (\alpha}\psi[A^{\mu )},A_{\beta}]
\partial_{\rho}h_{\alpha\mu}\}\n
&&+{1\over 2}Str exp(ik\cdot A)
[A^{\alpha},A^{\beta}][A^{\mu},A^{\nu}]\bar{\epsilon^1}\Gamma^{\rho} \psi
F_{\alpha\beta\mu\nu\rho} .
\eeqa
For simplicity we have neglected cubic terms in $\psi$ here.

In order to cancel the first line in the preceding expression which is
$O(1)$, we may modify the SUSY transformation eq.(\ref{Ssym1}) as follows:
\beqa
\delta^{(1)}\psi &=& \frac{i}{2}
                     [A_{\mu},A_{\nu}]\Gamma^{\mu\nu}\epsilon ^1
                     +ih_{\mu\alpha}P exp(ik\cdot A)
                     [A^{\mu},A_{\nu}]\Gamma^{\alpha\nu}\epsilon ^1 ,\n
\delta^{(1)} A_{\mu} &=& i\bar{\epsilon ^1}\Gamma_{\mu}\psi ,
\label{Ssym3}
\eeqa
where $Pexp(ik\cdot A) \hat{O}$ implies
$\int_0^1d\sigma exp(ik\sigma\cdot A)\hat{O}exp(ik(1-\sigma)\cdot A)$.
From SUSY transformation of the action, we obtain
\beqa
&&Str exp(ik\cdot A)
\{-i[A^{\alpha},\bar{\epsilon^1}\Gamma^{\beta}\psi][A^{\mu},A_{\beta}]
-{i\over 2}\bar{\epsilon^1}\Gamma^{\alpha\beta\gamma}\psi[A^{\mu},
[A_{\beta},A_{\gamma}]]\n
&&+\bar{\epsilon^1}\Gamma^{\alpha\beta\gamma}\psi
[k\cdot A,A_{\gamma}][A^{\mu},A_{\beta}]
-\bar{\epsilon^1}\Gamma^{\alpha}\psi
[k\cdot A,A^{\beta}][A^{\mu},A_{\beta}]
\}h_{\alpha\mu}.
\eeqa
Alternatively we can argue that
the above expression vanishes due to the equation of motion.
By combing these two contributions, we obtain
\beqa
&&-Str exp(ik\cdot A)
\{{i\over 2}\bar{\epsilon^2}\Gamma^{\alpha}\psi[k\cdot A,A^{\mu}]h_{\alpha\mu}
+\bar{\epsilon^2}\Gamma^{\rho\beta (\alpha}\psi[A^{\mu )},A_{\beta}]
\partial_{\rho}h_{\alpha\mu}\}\n
&&+Str exp(ik\cdot A)
\{\bar{\epsilon^1}\Gamma^{\alpha\beta\gamma}\psi
[k\cdot A,A_{\gamma}][A^{\mu},A_{\beta}]\n
&&+\bar{\epsilon^1}\Gamma^{\alpha}\psi[k\cdot A,A^{\beta}][A^{\mu},A_{\beta}]
-\bar{\epsilon^1}k\cdot\Gamma\psi[A^{\alpha}, A^{\beta}][A^{\mu},A_{\beta}]\n
&&-{1\over 4}\bar{\epsilon^1}\Gamma^{\alpha\beta\gamma}\psi
[A^{\mu},k\cdot A][A_{\beta},A_{\gamma}]
-{1\over 2}\bar{\epsilon^1}\Gamma^{\beta}\psi
[A^{\mu},k\cdot A][A_{\beta},A^{\alpha}]\}h_{\alpha\mu}\n
&&+{1\over 2}Str exp(ik\cdot A)
[A^{\alpha},A^{\beta}][A^{\mu},A^{\nu}]\bar{\epsilon^1}\Gamma^{\rho} \psi
F_{\alpha\beta\mu\nu\rho} .
\label{grvtrf}
\eeqa

We first focus on the terms which depend on $\epsilon^2$
in the first line of (\ref{grvtrf}).
These kind of terms also originate
from the gravitino coupling in eq.(\ref{grtnvt})
after postulating the local SUSY transformation of $\eta_{\mu}$ as
\beq
\tilde\delta \eta_{\mu}=-\partial_{\mu}\epsilon^2-
{1\over 2}\Gamma^{\nu\rho}\partial_{\nu}h_{\mu\rho}\epsilon^2 .
\label{grvstr0}
\eeq
The above expression can be obtained by
expanding the covariant derivative $D_{\mu}\epsilon^2$ in $h_{\alpha\mu}$
which is identified with zehnvein.
After this procedure however, we are still left with the following term
\beq
-{i\over 2} Str exp(ik\cdot A)
\bar{\psi}\Gamma^{\alpha\nu\rho}[A^{\mu},A_{\rho}]k_{\nu}\epsilon^2
h_{\alpha\mu} .
\label{apdx3}
\eeq

We may try to obtain the remaining $\epsilon^1$ dependent terms
from (\ref{antgrv}) by postulating the following
SUGRA transformation of anti-gravitino
which is complex conjugate to eq.(\ref{grvstr0})
\beq
\tilde\delta \bar{\eta}^c_{\mu}=\partial_{\mu}\bar{\epsilon^1}
-{1\over 2}\bar{\epsilon^1}\Gamma^{\nu\rho}\partial_{\nu}
h_{\mu\rho} .
\label{grvstr}
\eeq
Although we can cancel or account for half of the numerical coefficients
of a few terms in (\ref{grvtrf}), we also introduce new types of terms in
this procedure. We record the remaining discrepancy in what follows
\beqa
 &&{1\over 2} Str exp(ik\cdot A)
\{{1\over 2}\bar{\epsilon^1} k_{\nu}\Gamma^{\alpha\beta\gamma\nu\rho}\psi
[A^{\mu},A_{\alpha}][A_{\beta},A_{\gamma}]\n
&&+\bar{\epsilon^1} k_{\nu}\Gamma^{\nu\rho\beta}\psi
[A^{\mu},A_{\alpha}][A^{\alpha},A_{\beta}]
+\bar{\epsilon^1} k_{\nu}\Gamma^{\rho\beta\gamma}\psi[A_{\nu},A_{\gamma}]
[A^{\mu},A_{\beta}]\n
&&+\bar{\epsilon^1} k_{\nu}\Gamma^{\rho}\psi
[A_{\nu},A_{\beta}][A^{\mu},A^{\beta}]-\bar{\epsilon^1} k_{\nu}\Gamma^{\nu}\psi
[A^{\rho},A_{\beta}][A^{\mu},A^{\beta}]
\n
&&
-\bar{\epsilon^1} k_{\nu}\Gamma^{\beta}\psi
[A^{\mu},A_{\nu}][A_{\beta},A^{\rho}]
\}h_{\mu\rho} .
\label{apdx4}
\eeqa
The last term of eq.(\ref{grvtrf}) is not inconsistent with the
following SUGRA transformation of the anti-gravitino:
\beq
\tilde\delta \bar{\eta}^c_{\mu}=
{i\over 4\cdot 5!}\bar{\epsilon^1}\Gamma^{\mu}
\Gamma^{\rho_1\cdots\rho_5} F_{\rho_1\cdots\rho_5} .
\label{angrtr}
\eeq
Although we can account for one half of it in this manner,
we also obtain other terms with different tensor structures.

\end{document}